\documentclass[aps,prl,10pt,twocolumn, groupedaddress,floatfix,amsmath,amssymb]{revtex4-2}
\usepackage[normalem]{ulem}
\usepackage{graphicx}
\usepackage[english]{babel}
\usepackage[usenames,dvipsnames,pdftex]{color}
\usepackage[%
  colorlinks=true,
  urlcolor=blue,
  linkcolor=blue,
  citecolor=blue
]{hyperref}
\newcommand{\V}[1]{\boldsymbol{#1}}
\graphicspath{{Figures/}}
\usepackage{blindtext}
\usepackage[position=top, caption = false]{subfig}
\usepackage{pifont}
\usepackage[position=top]{subfig}
\usepackage{floatrow}
\usepackage{ulem}

\usepackage{colortbl}
\begin{document}
\title{Signatures of topological phase transitions in the s-wave superconductor\\at finite temperature}

\date{\today}

\author{Stefan Gorol}
\author{Florian Loder}
\author{Daniel Braak}
\author{Arno P. Kampf}
\author{Thilo Kopp}

\affiliation{Center for Electronic Correlations and Magnetism, Institute of Physics, 
University of Augsburg, 86135 Augsburg, Germany}

\begin{abstract}
In two dimensions, an s-wave superconductor in the presence of Rashba
spin-orbit coupling possesses distinct topologically
non-trivial ground state phases controlled by Zeeman splitting and
band filling. These phases can be characterized in terms of
spin textures in momentum space. Although the spin texture becomes
topologically trivial at finite temperatures, we identify thermodynamic
signatures that are directly related to the topological phase
transitions of the ground state. In particular, relative maxima in the
entropy as a function of the magnetic field in the vicinity of
topological phase transitions emerge and are attributed to a sign change
in the derivative of the magnetization with respect to temperature.
\end{abstract}
\maketitle
\section{Introduction}
Topologically non-trivial electronic systems in two dimensions (2d) are often characterized by a non-zero Chern number \cite{Bernevig2013,Berry,RevModPhys821959}  
which reflects a property of the ground state wave function.
This topological invariant is essential for the quantum Hall effect \cite{RevModPhys831057} and determines the quantized Hall conductivity in the zero temperature limit, or the thermal Hall effect \cite{thermalHall,PRB93, RevModPhys823045} for which the finite Chern number dictates the term in the thermal Hall conductivity linear in temperature. 
The 2d s-wave superconductor can be topological---in the sense that the Chern number may differ from zero---if spin-orbit coupling (SOC) and Zeeman spin splitting are present \cite{Sau,PhysRevB96024508, alicea, Zyuzin}.

In these systems, thermodynamic properties of
Lifshitz transitions concomitant with topological phase transitions have been investigated \cite{ThermodynEdge, ThermodynEdge2, ThermodynRashbaSC,spinExpectation}. Those transitions are
accompanied by
a kink in the first derivative of the thermodynamic potential with respect to the chemical potential at zero temperature \cite{LifshitzTransition,LifshitzTopology}. At finite temperatures, the kink is broadened
but may still be detected
as a peak in the third derivative of the thermodynamic potential with respect to the chemical potential
\cite{LifshitzFiniteT, ThermodynRashbaSC, ThermodynWeyl, ThermodynRashba}. 
On the other hand the bulk topology can be related to topological edge states by the bulk-boundary correspondence \cite{Bernevig2013,PRL49,Mong}. By using Hill thermodynamics the presence of the edge states in finite-size systems can be taken into account \cite{ThermodynEdge, ThermodynEdge2},
resulting in a linear contribution to the heat capacity.  

In addition to the Chern number, non-trivial bulk topology may be characterized by a non-zero skyrmion number for spin textures in reciprocal space \cite{skyrmion,PhysRevB96024508,upsala,Sau,QWZ}.
At zero temperature, the skyrmion number in a topological s-wave superconductor is related to the Chern number \cite{PhysRevB96024508}, but any finite temperature destroys its topological character as will be discussed in more detail in section~\ref{top}.
While the Lifshitz transition is related to the Fermi surface structure of the corresponding normal conducting phase, we study here thermodynamic signatures of the spin texture and predict a maximum of the entropy as function of magnetic field at constant temperature in the vicinity of the topological phase transition. This entropy maximum at non-zero temperatures may serve as an experimentally accessible way to detect the transition into a topologically non-trivial ground state of the system.

The realization of a topological 2d s-wave superconductor is, however, not straightforward, because the necessary magnetic fields---if applied perpendicular to the plane---are usually larger than the upper critical magnetic field $h_\mathrm{c2}$. It has been argued that topologically non-trivial phases in the s-wave superconductor may be accessible by magnetic field rotation towards in-plane orientations \cite{SciRep515302}, which preserves the non-trivial topology and avoids exceeding $h_\mathrm{c2}$ \cite{Beasly1975}. In this way, the topological phase transition may be observed in spin-orbit coupled s-wave superconductors with sufficiently strong
Zeeman splitting \cite{Sau}.

\begin{figure}
\begin{minipage}{\linewidth}
\subfloat[]{\includegraphics[width=1\linewidth]{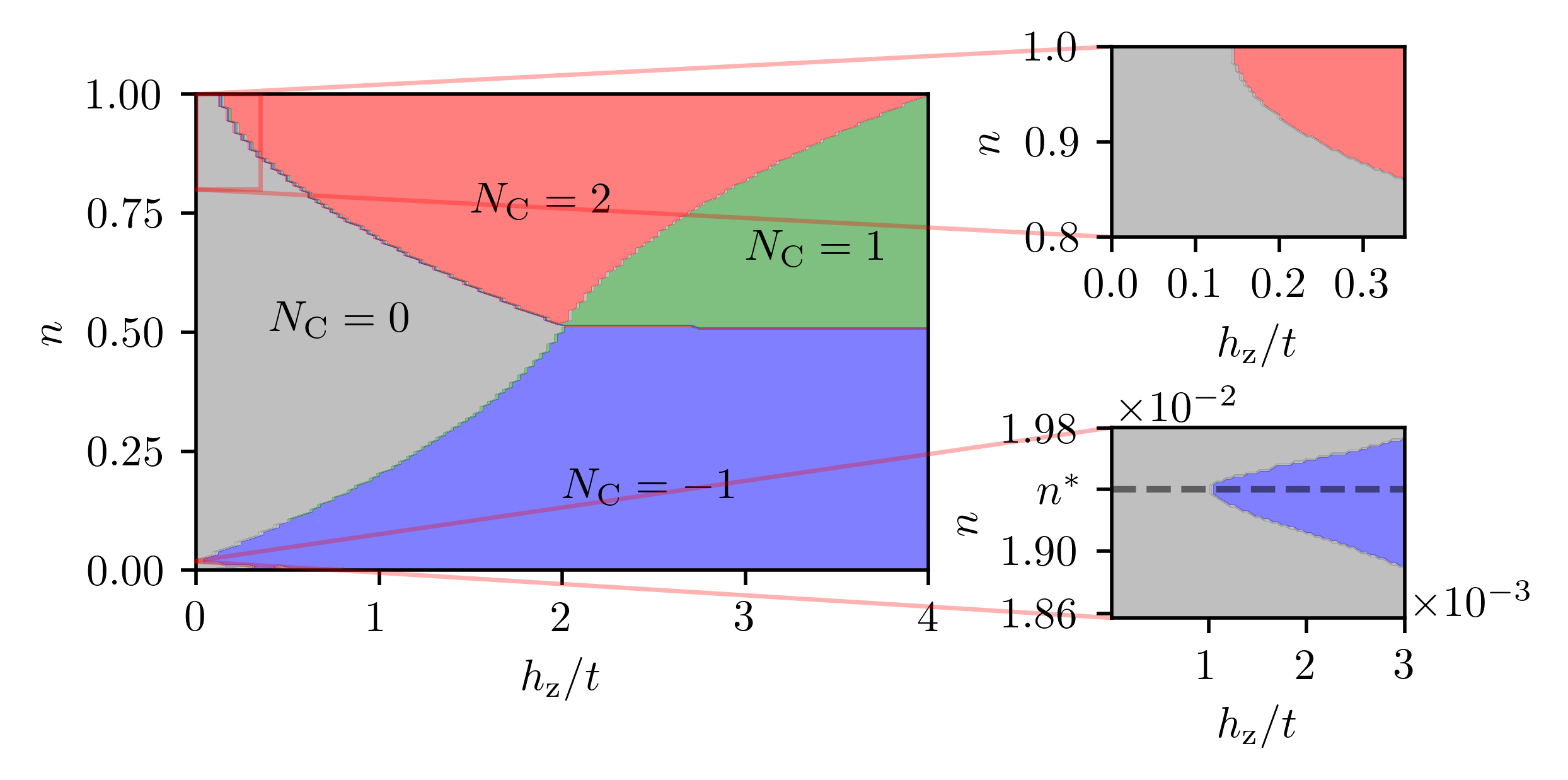}\label{phaseDiagram}}
\end{minipage}
\begin{minipage}{0.7\linewidth}
 \subfloat[]{\includegraphics[width=1\linewidth]{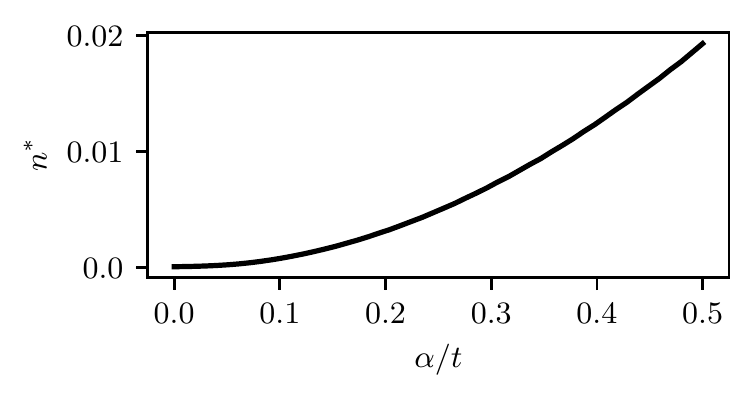}\label{n-star}}
 \end{minipage}
\caption{(a) Phase diagram of topological ground states specified by the Chern numbers $N_\mathrm C$ for $\alpha/t = 0.5$ and $V/t = 0.75$.
(b) Filling $n^*$ as a function of $\alpha/t$, defined as the band filling at which $h_{\mathrm t,1}=\Delta_\mathrm{OP}$ (see text).}
\label{phases}
\end{figure}

\section{Topological Characterization}\label{top}
First, we briefly review the model of an topological s-wave superconductor and its topological phases. We exclusively concentrate on the situation of intra-band pairing discussed in Reference\,\cite{JPhysCondensMatter25362201}.
The Bogoliubov-de Gennes Hamiltonian for the 2D superconducting system
under consideration is diagonal in momentum space and given by the 4x4 matrix
\begin{equation}
\mathcal{H}(\V k)=\left(\begin{array}{cc}
\mathcal H_0(\V k)&\mathbb D\\
\mathbb D^\dagger & -\mathcal H_0^\dagger(\V k)
\end{array}\right),
\label{Hamiltonian}
\end{equation}
where the Nambu spinor basis $\V \psi_{\V k} = \left(\hat c_{\V k,\uparrow},\hat c_{\V k,\downarrow},\hat c^\dagger_{-\V k,\uparrow},\hat c^\dagger_{-\V k,\downarrow}\right)^\dagger$ is used with $\V k = (k_\mathrm x,k_\mathrm y)^\mathrm T$.
The 2x2 matrix $\mathcal H_0$ reads 
\begin{equation}
\mathcal{H}_0=\left(\begin{array}{cc}\epsilon(\V k)-\mu+h_\mathrm z&\alpha(\V k)+h_\mathrm x-\mathrm i h_\mathrm y\\
{\alpha}^*(\V k)+h_\mathrm x+\mathrm i h_\mathrm y&\epsilon(\V k)-\mu-h_\mathrm z
\end{array}\right)
\end{equation}
while $\epsilon(\V k)=-2t\,(\cos k_\mathrm x +\cos k_\mathrm y)$ with $t$ and $\mu$ denoting the usual tight binding hopping energy and the chemical potential, respectively. The Rashba SOC is $\alpha(\V k)=\alpha\left( \sin k_\mathrm y+i\sin k_\mathrm x \right)$. The coupling of the magnetic field to the spin by the Zeeman term is included as $\mu_\mathrm{B}\V h\cdot\V \sigma$ where $h_\mathrm x$, $h_\mathrm y$ and $h_\mathrm z$ are the 
magnetic field components in x-,y- and z-direction, respectively. We use units such that $k_\mathrm B,\mu_\mathrm B,\hbar = 1$.
Regular s-wave pairing is assumed
\begin{equation}
\mathbb{D} =\left(\begin{array}{cc}
0&\Delta\\
-\Delta&0
\end{array}\right)
\end{equation}
with $\Delta$ being the order parameter. The grand canonical potential reads
\begin{equation}
\Omega = -\frac{1}{\beta}\sum_{\V k}\sum_\nu\ln\left[2\cosh\left(\frac{\beta \lambda_\nu(\V k,\mu)}{2}\right)\right]-2N\mu+\frac{N|\Delta|^2}{V}.
\end{equation}
where $\lambda_\nu$ denotes the eigenvalues of the Hamiltonian (\ref{Hamiltonian}) and $N$ is the number of lattice points. The attractive on-site interaction is $V>0$ and $\beta=1/T$. 
From the gap equation $\partial \Omega/\partial \Delta = 0$ the $\boldsymbol k$-independent mean-field order parameter $\Delta_\mathrm{OP}$ is obtained and
the particle number density is determined from $n = -\frac{1}{N}{\partial \Omega}/{\partial\mu}$. \\

In this system, a relevant topological invariant is the Chern number 
\begin{equation}
N_\mathrm C = \frac{1}{2\pi}\sum_{\nu,\mathrm{occupied}}\int\mathrm d^2 k\,\Omega^\nu_\mathrm B(\V k) ;\qquad N_\mathrm C \in \mathbb Z
\end{equation}
with the Berry curvature $\Omega^\nu_\mathrm B(\V k)$ given by
\begin{equation}
\Omega^\nu_\mathrm B(\V k) = \left.\mathrm i\nabla\times \langle \nu,\V k\vert \nabla_{\V k} \vert\nu,\V k\rangle\right|_z,
\end{equation}
where $\nu$ runs over all occupied bands and $\vert \nu,\V k\rangle$ refers to an eigenstate of the band $\nu$. For $\V h\ne \V 0$ the four eigenbands $\lambda_\nu$ with $\nu \in \{1,2,3,4\}$ are non-degenerate. The bands are in the following indexed by increasing energy. The Nambu-space bands are related by $\lambda_1(\V k,\mu) =-\lambda_4(\V k,\mu)$ and $\lambda_2(\V k,\mu)=-\lambda_3(\V k,\mu)$.

A further topological invariant is the skyrmion number $N_\mathrm S$, defined by
\begin{equation}
N_\mathrm S = \frac{1}{4\pi}\int d^2k\,\V S(\V k)\cdot\left(\partial_{k_x}\V S(\V k)\times \partial_{k_y}\V S(\V k)\right),
\end{equation}
where $\V S(\V k)$ is the normalized spin vector 
\begin{equation}
\V S(\V k) = \V s(\V k)/|\V s(\V k)|
\label{Snorm}
\end{equation}
with $\V s(\V k) = \left(\langle \hat s_x(\V k)\rangle , \langle \hat s_y(\V k)\rangle, \langle \hat s_z(\V k)\rangle\right)$.
We obtain the spin expectation values at each $\boldsymbol k$-point for a given temperature $T$ with 
\begin{widetext}
\begin{align}
\langle \V s(\boldsymbol k)\rangle &=\frac{1}{2}\sum_{\nu=1,2}\tanh\left(\frac{\beta\lambda_\nu(\boldsymbol k,\mu)}{2}\right)\langle \nu,\V k\vert \left(c^\dagger_{\V k,\uparrow}c^\dagger_{\V k,\downarrow}\right)\V \sigma \left(c_{\V k,\uparrow}c_{\V k,\downarrow}\right)^\mathrm{T}\vert \nu,\V k\rangle.
\end{align}
\end{widetext}
It was shown previously \cite{PhysRevB96024508} that in a system described by~(\ref{Hamiltonian}) the skyrmion number and the Chern number are related by $N_\mathrm C = -2N_\mathrm S$ at zero temperature. The skyrmion number takes half-integer values because the spin texture is actually of  meron character \cite{Nych,Woo,Guo}. In the following, for simplicity, we will omit expectation-value brackets and the hat identifying operators.\\

In order to enter any non-trivial topological phase it is necessary to apply magnetic field energies larger than $\Delta_\mathrm{OP}$ but the upper critical magnetic field energy set by $h^2_{\mathrm c2}$ is usually smaller than $\Delta_\mathrm{OP}$. 
If the magnetic field is rotated into an in-plane orientation, the topological transition fields $h_{\mathrm t,1}$ and $h_{\mathrm t,2}$ are decreased. 

For {$h_\mathrm x$, $h_\mathrm y\ll\alpha$}
\begin{equation}
h_{\mathrm t,1,2}(h_{\mathrm x,\mathrm y}\ne 0)\approx h_{\mathrm t,1,2}(h_{\mathrm x,\mathrm y}= 0)-t\frac{h_\mathrm x^2+h_\mathrm y^2}{\alpha^2}
\end{equation}
is obtained.
Concomitantly, orbital depairing is
suppressed in the rotated field setup such that $h_\mathrm{c,2}$ is increased in layered superconductors \cite{Beasly1975,WHH}. 
It may hence be possible to realize the situation where $h_{\mathrm{t},1}<h_{\mathrm{c,2}}$.
The required field rotation into an in-plane orientation leads to pairing with finite center-of-mass momentum \cite{PRL94137002,PRL89227002,PRL108117003,FFL,LO,PRL89227002,PRL108117003}, however, this does not destroy the inherent topological character of
the considered phases \cite{PhysRevB96024508} and, consequently, the finite temperature
signatures addressed in the following are not affected qualitatively. Therefore, without loss of generality, we discuss here the situation for $h_\mathrm x= h_\mathrm y = 0$. 
The topological phase-transition fields are defined by two magnetic fields $h_\mathrm{t,1}(n)$ and $h_\mathrm{t,2}(n)$, 
at which the band gap closes, 
given by \cite{Morr,PRL103020401}
\begin{align}
h_{\mathrm t,1}(h_{\mathrm x,\mathrm y}=0)&=\sqrt{\left({\Delta_\mathrm{OP}}\right)^2+(\epsilon(\V 0)-\mu(n))^2},\label{hts1}\\
h_{\mathrm t,2}(h_{\mathrm x,\mathrm y}=0)&=\sqrt{\left({\Delta_\mathrm{OP}}\right)^2+\mu^2(n)}.
\label{hts2}
\end{align}

\begin{figure}
\includegraphics[scale=0.25]{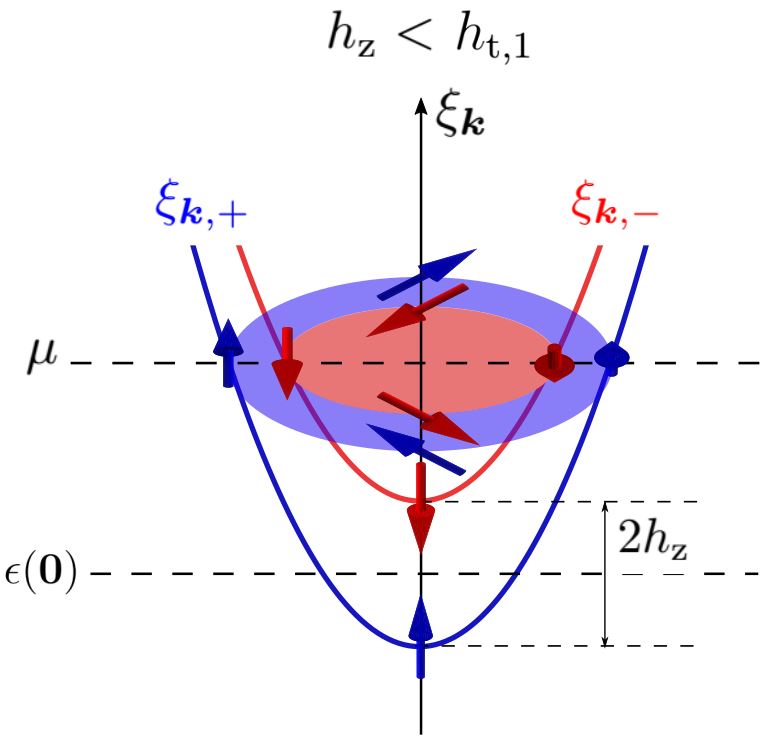}
\includegraphics[scale=0.25]{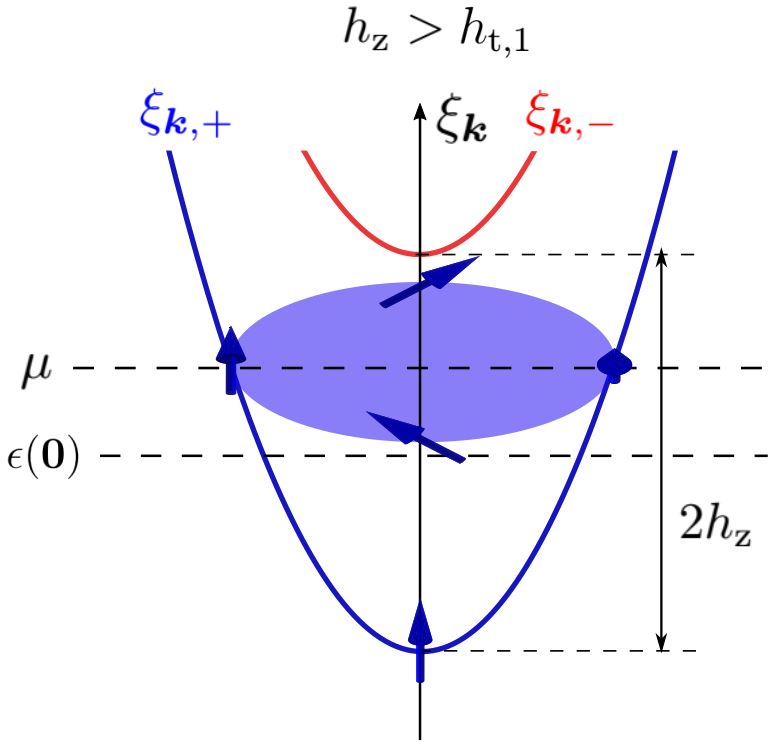}
\caption{Dispersion of the normal conducting bands $\xi_{\V k}^+$ and $\xi_{\V k}^-$ for the cases $h_\mathrm z<h_{\mathrm t,1}$ (left figure),
  $h_\mathrm z>h_{\mathrm t,1}$ (right figure). The arrows indicate the spin direction related to each band. The situation around $h_{\mathrm t,2}$ is similar.
}
\label{FermiLevel}
\end{figure}

The Chern number as a function of $h_\mathrm z$ and the band filling $n$ is calculated and the result is shown in Figure\,\ref{phases} 
indicating four topologically distinct phases with $N_{\mathrm C} \in\left\{0,-1,1,2\right\}$. 
Due to Rashba SOC, the superconducting order parameter is finite even for large magnetic fields,  if orbital depairing is not included, and it vanishes only in the $h_\mathrm z/t\to \infty$ limit \cite{physRevA92}.

For low band fillings $n\ll 1$ one has $\mu(n)\approx \epsilon(\V 0)$ and hence $h_{\mathrm t,1}<h_{\mathrm t,2}$. However, around half filling $n\approx 1$ one finds $\mu(n)\approx 0$ and thus $h_{\mathrm t,1}>h_{\mathrm t,2}$. In between these cases the crossing point of both transition-field curves is found at around quarter filling $n \approx 1/2$. At this point all topological phases merge as shown in Figure\,\ref{phaseDiagram}.

Gap closing points emerge whenever one band of the normal conducting state is depleted
--- related to $h_{\mathrm t,1}$ and  $h_{\mathrm t,2}$ of Equations (\ref{hts1}) and (\ref{hts2}) in the $\Delta_{\textrm{OP}} \to 0$ limit ---
as illustrated in Figure\,\ref{FermiLevel}. There, $\xi^+_{\V k}$ and $\xi_{\V k}^-$ are the helical spin-split eigenbands of $\mathcal H_0$.
The transition fields depend 
on  $\alpha$ through $\mu(n)$ obtained by solving the particle-number equation.

The non-trivial topological phase is characterized by $N_\mathrm C = -1$ in the low filling regime. Around half filling 
non-trivial topological phases with $N_\mathrm C = 2$ and $N_\mathrm C = 1$ are possible.
The latter requires large magnetic fields with $h_\mathrm z \ge \epsilon(\V 0)/2 = h_0$ 
where $ h_0$ is the minimal magnetic field for which $h_\mathrm z >h_{\mathrm t,1},h_{\mathrm t,2}$ is fulfilled.

\begin{figure}
\includegraphics[width=0.62\linewidth]{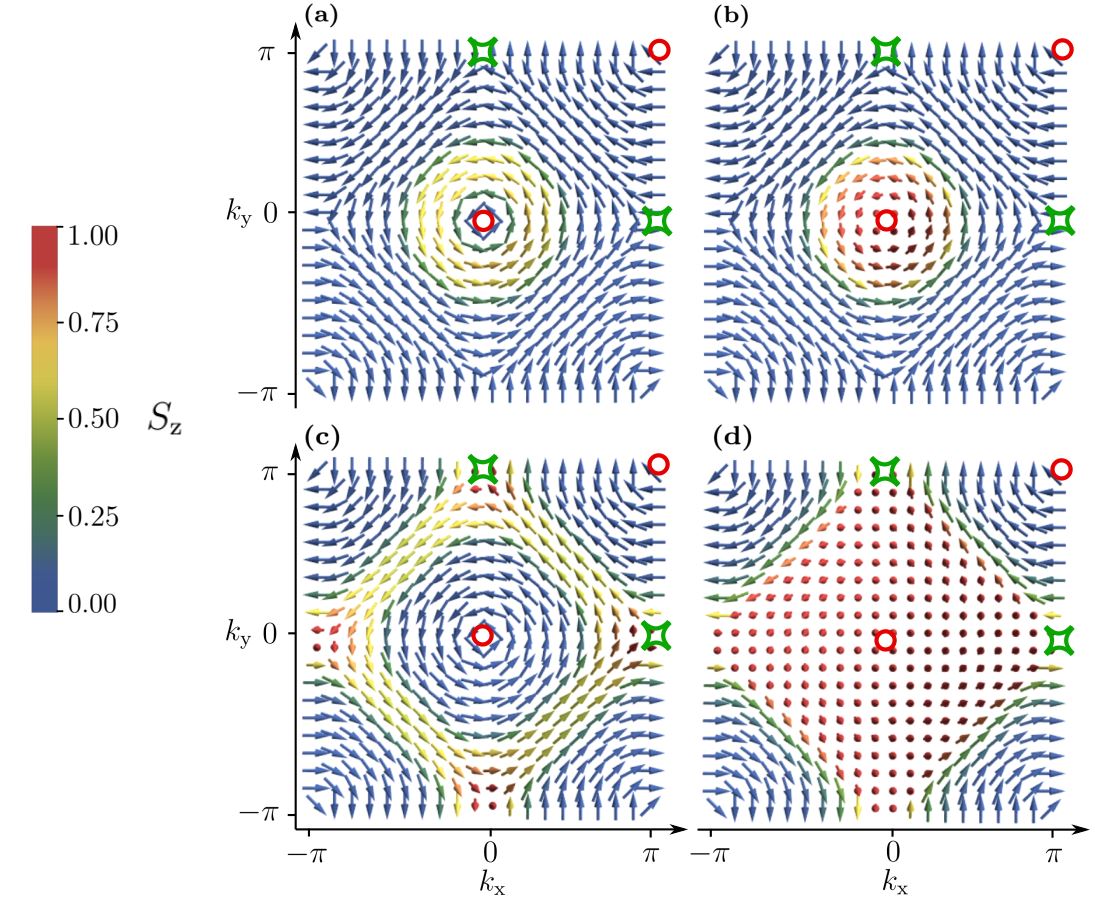}
\caption{Zero temperature spin textures for (a) the topologically trivial phase with $N_\mathrm S =N_\mathrm C= 0$, $n=0.19$, $(h_\mathrm z-h_{\mathrm t,1})/h_{\mathrm t,1}=-0.5$, (b) for the topologically non-trivial phases with $N_\mathrm S = 1/2$, $N_\mathrm C = -1$, $n=0.12$, (c) $(h_\mathrm z-h_{\mathrm t,1})/h_{\mathrm t,1}=0.18$,
  $N_\mathrm S = -1$, $N_\mathrm C = 2$, $n =0.9$, $(h_\mathrm z-h_{\mathrm t,2})/h_{\mathrm t,2}=0.3$ , and (d) $N_\mathrm S =- 1/2$, $N_\mathrm C = 1$, $n=0.54$, $(h_\mathrm z-h_{\mathrm t,1})/h_{\mathrm t,1}=0.05$. The arrows indicate the direction of the normalized spin vectors. The color codes the $z$-component of the normalized spin, where red corresponds to $S_\mathrm z = 1$ whereas blue corresponds to $S_\mathrm z = 0$. The positions of the $\V k_{\mathrm{VC},i}$, which are defined in the main text, are marked with red and green symbols for
vortices and antivortices, respectively.}
\label{SpinPhases}
\end{figure}

In the low filling regime, the minimum transition field $\min(h_{\mathrm t,1}(n))=\Delta_{\textrm{OP}}$ 
is obtained for a filling $n^*$ for which $\mu(n^*)=\epsilon(\V 0)$; $n^*$ depends on the Rashba SOC as shown in
Figure\,\ref{n-star}. Here it should be noted that $\epsilon(\V 0)$ is not the lowest band energy on account of the finite Rashba SOC. Since $N_\mathrm C = -2N_\mathrm S$ at zero temperature \cite{PhysRevB96024508}, the ground state phase diagram in Figure\,\ref{phaseDiagram} also amounts to four topologically distinct spin textures in reciprocal space.
At $h_{\mathrm{x},\mathrm{y}}=0$ and $T=0$, the spin texture for the topologically trivial phase for low band filling is depicted in 
Figure\,\ref{SpinPhases}\,a. 
There are vortex patterns at the time-reversal invariant momentum (TRIM) points $(0,0)$ and $(\pi,\pi)$ and antivortices at the TRIM points $(0,\pi)$ and $(\pi,0)$. The vortex (antivortex) center points are denoted as $\V k_\mathrm{VC}$ in the following. For $h_{\mathrm x,\mathrm y} \ne 0$, the $\V k_\mathrm{VC}$ are shifted away from the TRIM points.
At each center of a vortex or antivortex the spin normalization must be defined by the limit $\V k \rightarrow \V k_\mathrm{VC}$ which may be singular. For  $N_\mathrm C = 0$, it is singular at all four $\V k_\mathrm{VC}$. The normalized spin textures for the phases with $N_\mathrm C = -1,2,1$ are 
shown in 
Figures\,\ref{SpinPhases}\,b\,-\,d. 
In each of the these phases, the spin points upwards, $\V S=(0,0,1)^T$, at one or more of the  $\V k_\mathrm{VC}$ as depicted in Figure\,\ref{SpinPhases}.

As the normalized spin $\V S(\V k)$ can be viewed as a map from the torus (the 2d Brillouin zone) to the upper hemisphere of the sphere  S${}^2$, the number of the $\V k_\mathrm{VC}$ mapped onto the ``north pole'' are less then or equal to four, whereas the remaining points are mapped to the equator, where the map becomes singular. 
The manifold of the spin expectation values
can be compactified to the 
unit sphere such that the equator is mapped to the ``south pole'' of the sphere proving the spin texture's topological nature. In Figure\,\ref{SpinPhases}, the color of the arrows indicates on which 
latitude of the sphere the normalized spin expectation value is positioned
after compactification. Dark red corresponds to the covering of the north pole while dark blue implies the mapping onto the south pole.
Reference\,\cite{PhysRevB96024508} describes further details of this mapping. 

The number of north pole coverings is one, zero, two, and three for the phases $N_\mathrm C =-1,0,1,2$, respectively. At any finite temperature $T$, the $s_z$ spin components are finite everywhere in the Brillouin zone. The north pole is then covered four times since the in-plane spin components $s_\mathrm x=s_\mathrm y = 0$ by symmetry. Hence, the number ob $\V k_\mathrm{VC}$ mapped onto the equator is zero and the compactified map does not yield a full covering of the S${}^2$. Because the map is still smooth, the associated skyrmion number vanishes 
for any $T>0$.

\onecolumngrid

\twocolumngrid

In order to characterize the spin structure beyond the skyrmion number for finite temperature, the quantity 
\begin{equation}
\Sigma_\mathrm L = \frac{1}{2\pi}\int d^2 k\,\V S_\mathrm L (\V k)\cdot\left(\partial_{k_x}\V S_\mathrm L (\V k)\times \partial_{k_y}\V S_\mathrm L (\V k)\right),
\label{SigmaL}
\end{equation} 
was introduced in Reference\,\cite{PhysRevB96024508}
with the in-plane normalized spin vector
\begin{equation}
\V S_\mathrm L = \left(\frac{s_\mathrm x}{\sqrt{s_\mathrm x^2+s_\mathrm y^2}},\frac{s_\mathrm y}{\sqrt{s_\mathrm x^2+s_\mathrm y^2}},s_\mathrm z\right)=\left(S_\mathrm{L}^\mathrm x,S_\mathrm{L}^\mathrm y,s_\mathrm{z}\right).
\label{SL}
\end{equation}
$\Sigma_\mathrm L $ is not a topological invariant, and not confined to integer values. In fact, $\V S_\mathrm L(\V k)$ is a map from  the torus to the open unit cylinder, which is not compact. However, in the zero-temperature limit $\Sigma_\mathrm L \to N_\mathrm C$, because the map becomes smooth after the cylinder is compactified similar to the hemisphere above and $s_z$ is either zero or 1 at all vortex (antivortex) center points $\V k_\mathrm{VC}$.\\

By partial integration of (\ref{SigmaL}) we can rewrite $\Sigma_\mathrm L$ as
\begin{align}
\Sigma_\mathrm L &= \sum_{i=1}^4s_\mathrm z(\V k_{\mathrm{VC},i})\lim_{\epsilon\to0}\hspace{-0.5cm}\oint\limits_{C(\V k_{\mathrm{VC},i}, \epsilon)}^{} \hspace{-0.5cm}\frac{\mathrm d\V k}{2\pi}\,\left(S_\mathrm{L}^\mathrm x\nabla_{\V k}S_\mathrm{L}^\mathrm y-S_\mathrm{L}^\mathrm y\nabla_{\V k}S_\mathrm{L}^\mathrm x\right) \notag\\&=\sum_{i=1}^4s_\mathrm z (\V k_{\mathrm{VC},i})\mathcal V(\V k_{\mathrm{VC},i}).\label{SigmaLfinal}
\end{align}
The index $i$ runs over all vortex-center points where $\langle s_\mathrm x(\V k_{\mathrm{VC},i}) \rangle = \langle s_\mathrm y(\V k_{\mathrm{VC},i}) \rangle=0 $, and the transverse components of $\V S_\mathrm L$ cannot be normalized. The quantity $\mathcal V(\V k)$ denotes the winding number (vorticity) of the in-plane spin around the momentum $\V k$. The spin winding results from Rashba SOC.
Whereas the number of points $\V k_\mathrm{VC}$ with $\V S(\V k_\mathrm{VC}) =\V 0$ changes at the topological phase transition, 
their vortex or anti-vortex character is preserved. $C(\V k_{\mathrm{VC},i} ,\epsilon)$ denotes a counterclockwise circular path around $\V k_{\mathrm{VC},i}$ with radius $\epsilon$. $\Sigma_\mathrm{L}$ is integer valued, if $s_\mathrm{z}$ takes the extremal values 0 or 1 at the vortex points, as is the case for $T=0$.  
As all $\mathcal V(\V k_{\mathrm{VC},i})$ are unchanged at the topological phase transition 
$\Sigma_\mathrm L$ is exclusively determined by the values of $s_\mathrm z(\V k_{\mathrm{VC},i})$ and these change only at the gap closing points. 

\begin{widetext}

\begin{figure}
\begin{minipage}{0.49\linewidth}
\raggedleft
 \subfloat[]{\includegraphics[width=0.8\columnwidth]{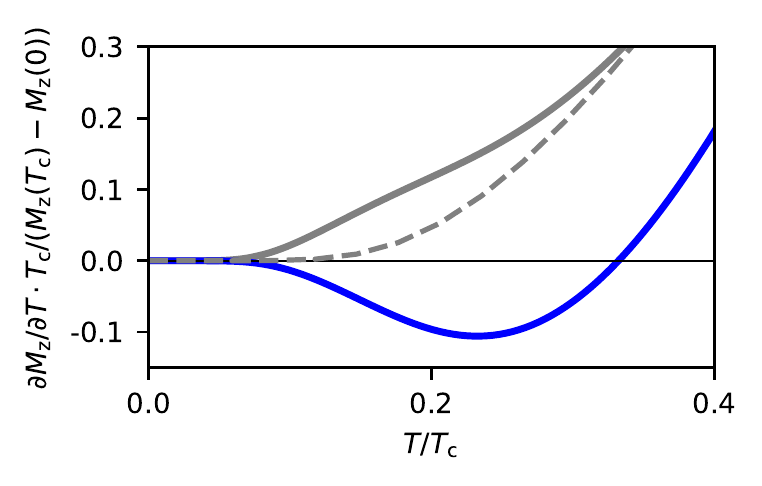}\label{fig:dMzdT}}\\
 \vspace{-1cm} \subfloat[]{\includegraphics[width=0.8\columnwidth]{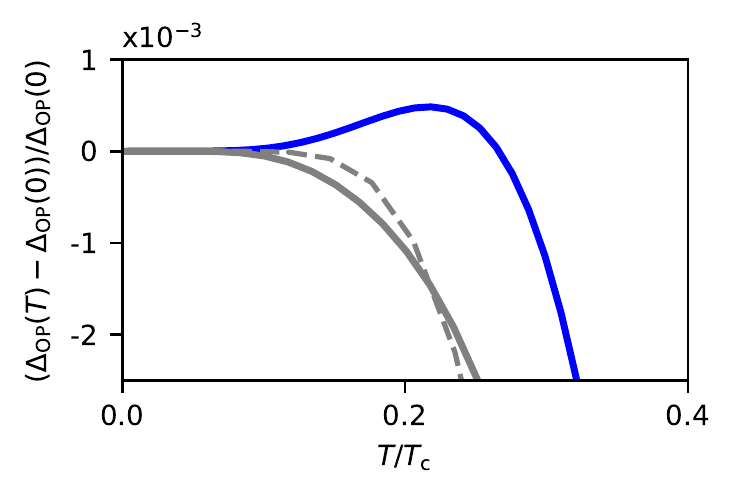}\label{fig:Delta_BCS}}
\end{minipage}
\begin{minipage}{0.49\linewidth} 
\raggedright
\subfloat[]{\includegraphics[width=0.8\columnwidth]{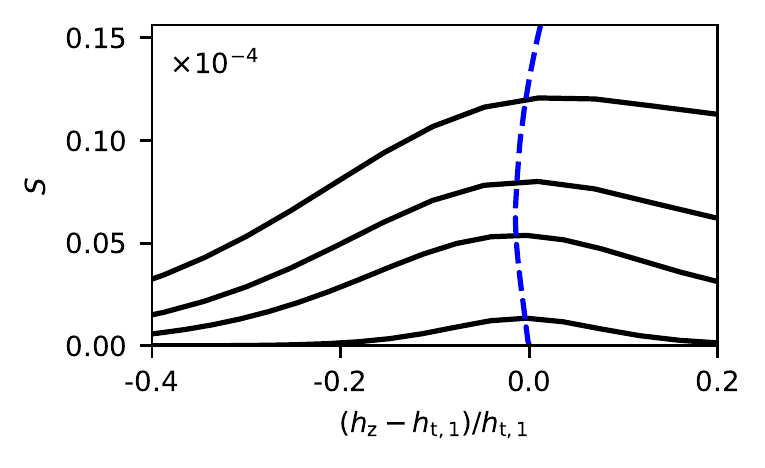}\label{fig:entropy}} \\[-0.5cm]
 \subfloat[]{\includegraphics[width=0.8\columnwidth]{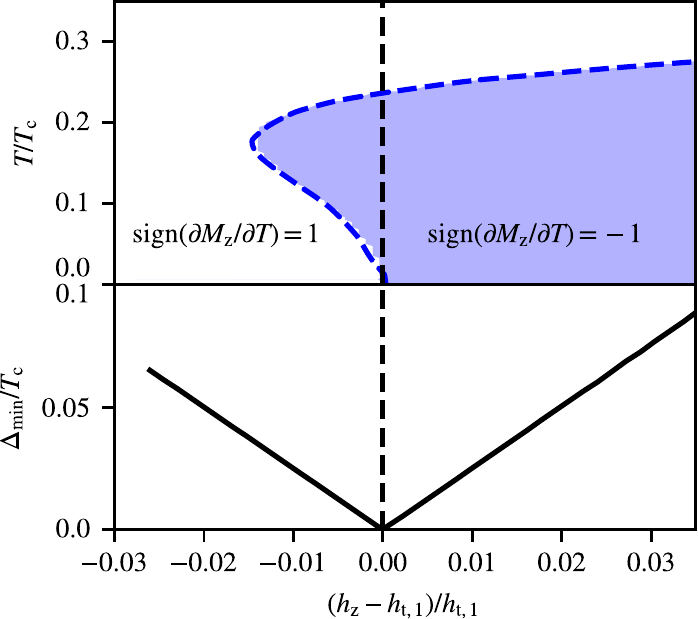}\label{fig:dMdTphase}} 
 \end{minipage}
\caption{Thermodynamic signatures of topological spin-texture changes in the ground state close to the transition field $h_{\mathrm t,1}$. (a) $\partial_T M_\mathrm z \cdot T_\mathrm c/(M_{\mathrm z}(T_\mathrm c)-M_\mathrm z(0))$ while $(M_{\mathrm z}(T_\mathrm c)-M_\mathrm z(0))$ is always positive and (b) $(\Delta_\mathrm{OP}(T)-\Delta_\mathrm{OP}(0))/\Delta_\mathrm{OP}(0)$ as functions of $T/T_\mathrm c$ for $(h_\mathrm z-h_{\mathrm t,1})/h_{\mathrm t,1}=0.19$, $\Delta_\mathrm{OP}(T=0)=0.05t$ (blue) and $(h_\mathrm z-h_{\mathrm t,1})/h_{\mathrm t,1}=-0.19$, $\Delta_\mathrm{OP}(T=0)=0.06t$ (grey); s-wave superconductor without SOC for $\Delta_\mathrm{OP}=0.05t$ and $h_\mathrm z = 0.6\Delta_\mathrm{OP}$ (dashed grey). (c) Entropy $S(h_\mathrm z)$ for $T/T_\mathrm c = 0.11,0.21,0.23,0.27$ from bottom to top. The blue dashed line connects the local maxima of $S(h_\mathrm z,T)$.
(d) $\partial M_\mathrm z/\partial T <0$ (blue area); $\partial M_\mathrm z/\partial T>0$ (white area) and minimum gap $\Delta_\mathrm{min}$ as a function of $(h_\mathrm z-h_{\mathrm t,1})/h_{\mathrm t,1}$ (solid black). We used $V/t = 1.5$ corresponding to a $T_\mathrm c \approx 3\cdot 10^{-2}t$, $n = 0.028$ and $\alpha/t = 0.5$. 
}
\label{dMdT}
\end{figure}
\end{widetext}

\section{Thermodynamics}

The values of $s_\mathrm z(\V k_{\mathrm{VC},i})$ at isolated points of the BZ 
are not accessible by measurements of thermodynamic quantities. 
However, the modification of the spin-structure in a sizable region of the Brillouin zone can affect thermodynamic properties and thereby exhibit signatures of the topological ground-state transition even at
non-zero temperatures. 

The energy gap $\Delta_\mathrm{min}$ depends on $\V k$ due to the Rashba SOC which is discussed in more detail in Reference\,\cite{JPhysCondensMatter25362201}. At low temperatures $T\ll \Delta_\mathrm{min}$ thermal excitations are restricted to a small neighborhood of the energy gap minima. Thus, information about 
variations in the $s_\mathrm z(\V k)$-values in the vicinity of the gap closing points can be obtained. Taking the partial derivative of the magnetization in z-direction with respect to 
temperature, and neglecting 
contributions further away from the Fermi-level yields
\begin{widetext}
\begin{equation}
- \left.\frac{\partial}{\partial T}\left(\left.\frac{\partial \Omega}{\partial h_\mathrm z}\right|_{T,\mu}\right)\right|_{h_\mathrm z,n}=\left.\frac{\partial M_\mathrm z}{\partial T}\right|_{n,h_\mathrm z}\hspace{-0.5cm}=\left.\frac{\partial S}{\partial h_\mathrm z}\right|_{n,T} \hspace{-0.5cm}= \frac{1}{N} \!\!\int\mathrm d^2 k\,\frac{\partial \lambda_3(\V k,\mu)}{\partial h_\mathrm z} \frac{\lambda_3(\V k,\mu)}{T^2}\;\mathrm{sech}^2\left(\frac{\lambda_3(\V k,\mu)}{2T}\right).
 \label{eq:thermodynamic}
\end{equation} 
\end{widetext}
We omitted the contributions of two eigenbands $\lambda_1$ and $\lambda_4$ which are far away from the Fermi level and used the symmetry $\lambda_2(\V k,\mu)=-\lambda_3(\V k,\mu)$.
$T\ll \Delta_\mathrm{min}$ is reflected in a peak in $T^{-2}\mathrm{sech}^2(\lambda_3(\V k,\mu)/2T)$ at the gap minimum $\V k_\mathrm{min}$ where $\lambda_3(\V k,\mu)$ is minimal. 
In Equation\,(\ref{eq:thermodynamic}), $S$ is the entropy and a Maxwell relation is used to relate the derivate of the magnetization with respect to $T$ to the derivative of the entropy with respect to the magnetic field.

In the following, we discuss the thermodynamic signatures 
of a topological ground-state phase transition
exemplarily in the low filling regime. 
Figure\,\ref{fig:dMzdT} shows $\partial_T M_\mathrm z$ for $h_\mathrm{x}=h_\mathrm y = 0$ and $h_\mathrm z$ above and below $h_{t,1}$, see blue and grey line, respectively. For magnetic fields below the topological phase transition $\partial_T M_{\mathrm z}>0$ for $T<T_\mathrm c$ similar to the case of an s-wave superconductor without SOC except in a narrow region around the topological ground state transition field $h_{\mathrm{t},1}$ as shown in Figure\,\ref{fig:dMdTphase}. 
In contrast, the figure shows $\partial_T M_{\mathrm z}$ being non-monotonous and even negative as a function of $T$ for $h_\mathrm z>h_{\mathrm t,1}$ and $T\ll T_\mathrm c$. The temperature scale for non-vanishingly small $\partial_T M_\mathrm z$ is set by the band gap minimum which is proportional to $h_\mathrm z-h_{\mathrm t,1}$ as shown in Figure\,\ref{fig:dMdTphase}. The dependence of $\partial_T M_\mathrm z$ on $T$ for a superconductor without SOC is given by the dashed grey curve where $\partial_T M_\mathrm z$ is exponentially increasing with $\Delta_\mathrm{OP}$. 
The temperature dependence of the order parameter $\Delta_\mathrm{OP}$ is qualitatively similar to the dependence of $\partial_TM_\mathrm z$ as depicted in Figure,\,\ref{fig:Delta_BCS}.

At $T = 0$, $s_\mathrm z(\V 0)=0$ for magnetic fields $h_\mathrm z <h_{\mathrm t,1}$ 
because the spin expectation values of the chiral spin split bands compensate each other. However, $s_\mathrm z$ turns into a maximum if $h_\mathrm z$ is tuned through 
the topological phase transition
(the image in the spin map of the momentum $\V k=\V 0$ switches from the south to the north pole). 
Finite temperature excitations now have a quantitatively different effect on the $s_\mathrm z(\V k = 0)$-values in the trivial and the topological phase where $s_\mathrm z(\V k = 0, T)$ is increasing and decreasing, respectively. 

This qualitative difference extends to a finite region around the $\Gamma$-point such that the change from the minimum in $s_\mathrm z$ to the maximum is visible as a sign change in $\partial_T M_{\mathrm z}$. 
At $T\ll \Delta_\mathrm{min}$ the integral in (\ref{eq:thermodynamic}) can be approximated by the value of the integrand at the band gap minimum due to the sharp peak in $T^{-2}\mathrm{sech}^2(\lambda_3/2T)$. The sign of $\partial_T M_\mathrm z$ is therefore determined by the sign of $\partial_{h_\mathrm z}\lambda_3$ at the band gap minimum. 

At finite temperatures in the regime $T\ll T_\mathrm c$, the sign change of $\partial_T M_\mathrm z$ does not occur exactly at the transition field $h_{\mathrm t,1}$ but 
in a magnetic field range where the two normal conducting bands are 
still filled in the vicinity of  the $\Gamma$-point. 
The points at which $\partial_T M_\mathrm z=0$ (as a function of $h_\mathrm z$ and $T$) are given by the blue dashed line in Figure\,\ref{fig:dMdTphase}. They correspond
to the positions of the relative maxima in the entropy in the region with $T\ll \Delta_\mathrm{min}$ around $h_{\mathrm t,1}$ as shown by the blue dashed line in Figure\,\ref{fig:entropy}.

\section{Conclusion and final remarks}
Topologically non-trivial states can
be realized in s-wave superconductors in the presence of Zeeman splitting
and Rashba spin-orbit coupling \cite{SciRep515302}.
The topologically non-trivial phases in the system may be described by either the Chern number or the skyrmion number for the spin-texture in momentum space. The Chern number description is valid only for the ground state while the skyrmion number 
characterization yields trivial topology at any finite temperature. Nevertheless, there exist signatures of the topological phase transition at finite temperature related to the profound change in the spin texture signified in certain thermodynamic quantities. 

A topological superconductor with spin-orbit coupling shows a characteristic vortex structure of the in-plane spin components at the time-reversal invariant momenta in the Brillouin zone. The out-of-plane spin component at distinct vortex centers flips from zero to one at a critical Zeeman splitting for $T=0$ and attains a maximum in the topologically non-trivial phase even for $T>0$. This sudden change manifests itself in a maximum of the entropy as function of magnetic field at constant temperature. Equivalently, it corresponds to a sign change of the derivative of the magnetization with respect to temperature. We propose a suitable generalization of the skyrmion number, $\Sigma_\mathrm L$,  which measures the singular character of the spin texture and is not confined to the ground state. In this way, $\Sigma_\mathrm L$ captures the topological nature of the phase transition even at non-zero temperature although it is not a topological invariant. 

Other two-dimensional systems with broken time-reversal symmetry, whose topological ground state depends on the magnetic field e.\,g. the topological mixed-parity superconductor \cite{DplusP1, DplusP2}, are candidates for similar investigations close to topological phase transitions. 
Thermodynamic signatures of topological phase transitions are not restricted to topological superconductors but expected as well for topological insulators.
Replacing the magnetization by a pseudospin polarization, the Chern insulators described in terms of topologically non-trivial pseudospin textures may be analyzed at finite temperatures in a similar fashion \cite{QWZ}.

\medskip
\begin{acknowledgments}
\textbf{Acknowledgements}\\
Financial support by the Deutsche Forschungsgemeinschaft (project number 107745057, TRR 80) is gratefully acknowledged.
\end{acknowledgments}
\bibliographystyle{apsrev4-2}
\bibliography{References}
\cleardoublepage

\end{document}